\documentclass[aps,pre,preprint,superscriptaddress,nofootinbib,longbibliography]{revtex4-2}

\usepackage{amsmath,amssymb,bm}
\usepackage{graphicx}
\usepackage{booktabs}
\usepackage{microtype}
\usepackage{hyperref}
\hypersetup{hidelinks}

\newcommand{\dd}{\mathrm{d}}
\newcommand{\ii}{\mathrm{i}}
\newcommand{\kb}{k_{\mathrm B}}
\newcommand{\eps}{\varepsilon}
\newcommand{\Pinf}{P_{\infty}}
\newcommand{\Ptot}{P_{\mathrm{tot}}}
\newcommand{\Peq}{P_{\mathrm{eq}}}
\newcommand{\PPW}{\Delta P_{\mathrm{PW}}}
\newcommand{\PEW}{\Delta P_{\mathrm{EW}}}

\newcommand{\appsection}[1]{\section{\MakeUppercase{#1}}} 
\makeatletter 
\let\old@abstractheading\frontmatter@abstractheading
\renewcommand{\frontmatter@abstractheading}{\vspace{1cm}\old@abstractheading}
\makeatother

\begin{document}

\title{Passive-reflector bounds and robust bilateral nonequilibrium Casimir repulsion}

\author{Mohamed R. Khodja}
\email{mkhodja@pmu.edu.sa}
\affiliation{Department of Electrical Engineering, Prince Mohammad Bin Fahd University, Khobar 31952, Saudi Arabia}

\author{Belkacem E. Khodja}
\affiliation{Department of Physics, Boston University, 590 Commonwealth Avenue, Boston, Massachusetts 02215, USA}

\begin{abstract}
Nonequilibrium electromagnetic fluctuations can generate repulsive Casimir--Lifshitz pressures, but optimizing a cavity at one operating point does not establish that the repulsion survives uncertainty or exceeds ordinary far-field radiation pressure. We formulate a two-sided biased-semiconductor cavity as a constrained max--min problem that maximizes the smaller outward pressure on its two reflectors. In the optically thick-slab limit, identical reflectors admit a symmetric global optimizer, and a passive-reflector bound separates cavity enhancement from the driven far-field plateau. At $300~\mathrm{K}$, a biased-GaAs cavity with $\pm10~\mathrm{nm}$ gap uncertainty and $\pm0.005$ drive uncertainty sustains a worst-case pressure of $2.257~\mathrm{mPa}$ with ideal reflectors. Reoptimization with a causal gold response retains $2.164~\mathrm{mPa}$. The nominal and robust optima lie within $2~\mathrm{nm}$ of one another, so the main value of the max--min calculation here is a quantitative performance certificate rather than a large geometric redesign. A full finite-slab calculation and an alternative published GaAs response change the reported pressures by at most $1.60\times10^{-3}\%$ and $4.44\times10^{-2}\%$, respectively. These results establish a reproducible design bound for bilateral nonequilibrium repulsion under dimensional, drive, reflector, and optical-model variations.
\end{abstract}

\maketitle

\newpage 

\section{Introduction}

The Casimir interaction between neutral macroscopic bodies is a direct consequence of electromagnetic fluctuations constrained by material boundaries \cite{Casimir1948,Lifshitz1956,DLP1961}. Causality and passivity also impose useful bounds on the equilibrium force between partially transmitting mirrors \cite{LambrechtJaekelReynaud1997}. For two parallel perfect electric conductors (PECs) separated by a vacuum gap $d$ at zero temperature, the pressure is attractive,
\begin{equation}
P_{\mathrm{PEC}}(d)=-\frac{\pi^2\hbar c}{240d^4}.
\label{eq:pec}
\end{equation}
For realistic materials the Lifshitz formulation replaces ideal mode counting by the causal, dispersive response of the bodies. At equilibrium, that response enters primarily through analytic continuation to imaginary frequency. Consequently, a striking real-frequency optical feature, including a negative-index band, does not by itself imply an equilibrium repulsive force \cite{Engheta2002,LeonhardtPhilbin2007,RosaDalvitMilonni2008,Klimchitskaya2009}. Repulsion is possible in special equilibrium arrangements, for example electric--magnetic idealizations \cite{Boyer1974}, suitable intervening liquids \cite{Munday2009}, or geometries and material classes outside the assumptions of attraction theorems \cite{KennethKlich2006,Rahi2009}.

A qualitatively different route is to drive the electromagnetic field out of equilibrium. Temperature differences, external pumping, and nonzero photon chemical potential alter source occupations at real frequency and therefore add force contributions that are not derivatives of an equilibrium free energy \cite{Antezza2008,MessinaAntezza2011,Kruger2011,Kruger2012,Bimonte2017}. Chen and Fan showed that a nonzero photon chemical potential can strongly modify fluctuation forces \cite{ChenFan2016}. More recently, Spreng \emph{et al.} demonstrated that moderately forward-biased direct-gap semiconductors can produce experimentally relevant repulsive pressures against metallic reflectors; the propagating contribution exhibits Fabry--P\'erot oscillations with separation and may overwhelm the attractive equilibrium term \cite{Spreng2024}. Control protocols for thermal nonequilibrium Casimir forces have also been developed, including feedback stabilization and trajectory tracking in multiplate systems \cite{IizukaFan2021,IizukaFan2022}. A recent review places biased-semiconductor schemes among the principal proposed routes to repulsive Casimir forces in nanotechnology \cite{Shelden2024}.

The natural question is therefore no longer whether one selected bias, material, and separation happens to yield a positive pressure. It is a design problem: \emph{what physically admissible geometry and drive maximize a repulsive pressure that remains positive under admitted uncertainty and is demonstrably associated with cavity filtering rather than merely with the separation-independent far-field emission plateau?} This question is close in spirit to inverse-design work in macroscopic quantum electrodynamics \cite{BennettBuhmann2020} and recent algorithmic searches for Casimir--Polder repulsion \cite{Kilianski2025}, but the design variables and constraints are different. Existing biased-semiconductor calculations map force versus separation, band gap, and bias for prescribed material pairs \cite{Spreng2024}; the control literature prescribes a time-dependent force or trajectory \cite{IizukaFan2021,IizukaFan2022}. Neither addresses the static bilateral max--min design considered here, in which both boundary forces, a continuous uncertainty set, and the far-field radiation background enter the feasibility test.

The present work makes three main contributions. First, the two vacuum gaps are treated independently and the objective is the worst outward pressure on \emph{both} reflectors. This distinction matters out of equilibrium because forces on different bodies need not be equal and opposite \cite{Kruger2011,Kruger2012}. For identical boundaries we prove that the reduced bilateral problem admits a symmetric global optimizer, resolve its continuous inner uncertainty minimization by combining an occupation-monotonicity condition with a stationary-point audit, and derive the quadratic gap-uncertainty and linear drive-uncertainty penalties. Second, we derive the driven, phase-averaged far-field plateau for an arbitrary passive planar reflector. This result is distinct from equilibrium passive-mirror force bounds \cite{LambrechtJaekelReynaud1997}: a modewise inequality bounds the nonequilibrium radiation-pressure background, while a reflector-specific cavity-contrast constraint prevents the optimizer from selecting that background as a cavity optimum. Third, we test what the optimized result actually depends on. Nominal and robust designs are cross-evaluated, the uncertainty widths and required cavity excess are varied, and the calculation is repeated with finite-conductivity gold and an alternative published GaAs response. The band-gap screen, slab-thickness criteria, and full finite-slab calculation provide supporting consistency checks.

The primary numerical platform is forward-biased GaAs at 300 K. PEC reflectors provide the analytically transparent reference problem, and identical semi-infinite gold reflectors described by a causal Lorentz--Drude fit provide the nonideal benchmark. The calculation remains reciprocal and planar. This keeps the added test tied directly to the mechanism already studied, rather than introducing tensor response parameters that are not needed to decide whether the robust repulsion is a PEC artifact.

\section{Geometry and force definition}
\label{sec:geometry}

\begin{figure}[t]
\includegraphics[width=0.9\columnwidth]{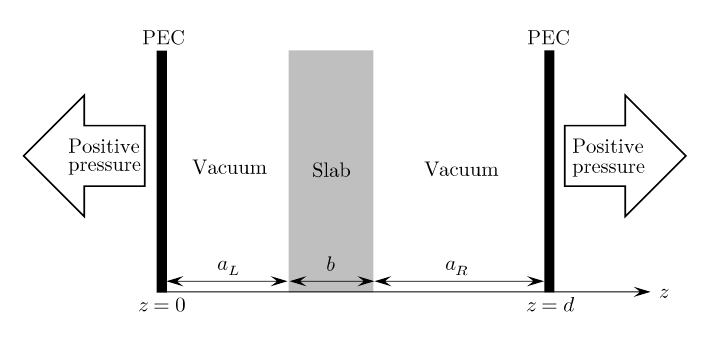}
\caption{Reflector--vacuum--biased-semiconductor--vacuum--reflector geometry and sign convention. $P_L>0$ and $P_R>0$ denote forces pushing the left and right reflectors outward, respectively. The gaps $a_L$ and $a_R$ are optimized independently; the semiconductor thickness $b$ is then selected by the reduction criteria of Sec.~\ref{sec:thickness}.}
\label{fig:geometry}
\end{figure}

Figure~\ref{fig:geometry} shows two infinite, identical planar reflectors normal to the $z$ axis. A semiconductor slab of thickness $b$ lies between them, leaving vacuum gaps $a_L$ and $a_R$. The slab is maintained in a nonequilibrium steady state by forward bias. The reflectors and the lattice of the semiconductor are taken to be at $T=300$ K; the nonequilibrium drive enters through the photon chemical potential rather than through a temperature difference. PECs are used for the reference calculation, followed by a reoptimization with gold.

For a sufficiently opaque central slab, the fields emitted through its two faces are exponentially decoupled. In that controlled limit the outward pressure on each PEC is a one-gap function,
\begin{equation}
P_L=P(a_L,\eta;\mathcal R),\qquad
P_R=P(a_R,\eta;\mathcal R),
\label{eq:bilateral}
\end{equation}
where
\begin{equation}
\eta\equiv\frac{\mu_\gamma}{E_g},
\label{eq:eta}
\end{equation}
$E_g$ is the semiconductor band gap and $\mu_\gamma$ is the photon chemical potential generated by the quasi-Fermi-level splitting. In an ideal forward-biased direct-gap semiconductor $\mu_\gamma\simeq eV_b$ over the luminescent band \cite{Wurfel1982,ChenFan2016,Spreng2024}. The parameter $\eta$ is treated here as a spatially uniform nonequilibrium-reservoir variable. Physically, it represents an approximately uniform quasi-Fermi-level splitting over the optically active regions of the two semiconductor faces. The model does not solve a carrier-transport problem and does not assert that an arbitrary electrically contacted slab realizes this condition. Allowing $\eta$ to vary with depth would require local source correlations coupled to drift--diffusion and recombination equations and constitutes a different optimization problem.

Here $\mathcal R$ denotes the common reflector model. Equation~\eqref{eq:bilateral} does not assert equilibrium action--reaction symmetry. Rather, it is the local stress on each reflector. A nonequilibrium steady state can exchange momentum with the pump, the semiconductor support, and the electromagnetic field. The design objective below therefore requires both $P_L$ and $P_R$ to be positive separately. In the symmetric optimum, mirror symmetry additionally removes a net $z$-directed thrust on the central slab within the ideal model.

\section{One-gap fluctuation pressure}
\label{sec:pressure}

We first derive the pressure between one semi-infinite biased semiconductor and one passive planar reflector separated by a vacuum gap $a$. The total outward pressure on the reflector is written
\begin{equation}
P(a,\eta;\mathcal R)=\Peq(a;\mathcal R)+\PPW(a,\eta;\mathcal R)+\PEW(a,\eta;\mathcal R),
\label{eq:totalpressure}
\end{equation}
where the first term is the equilibrium Lifshitz pressure at the common temperature $T$, and the last two terms are the excess nonequilibrium contributions from propagating and evanescent modes. Positive pressure is repulsive. The reflector argument is suppressed below when no ambiguity can arise.

\subsection{Reflection amplitudes}

For an isotropic semiconductor with dielectric function $\eps(\omega)$, define $k=|\bm{k}_\parallel|$ and
\begin{equation}
k_z=\sqrt{\frac{\omega^2}{c^2}-k^2},\qquad
k_{zs}=\sqrt{\eps(\omega)\frac{\omega^2}{c^2}-k^2},
\end{equation}
with branches chosen so that $\operatorname{Im}k_z\ge0$ and $\operatorname{Im}k_{zs}\ge0$. The vacuum-to-semiconductor Fresnel amplitudes are
\begin{align}
r_s^{\mathrm{TE}}&=\frac{k_z-k_{zs}}{k_z+k_{zs}},\\
r_s^{\mathrm{TM}}&=\frac{\eps k_z-k_{zs}}{\eps k_z+k_{zs}}.
\label{eq:fresnel}
\end{align}
For the PEC reference,
\begin{equation}
r_c^{\mathrm{TE}}=-1,\qquad r_c^{\mathrm{TM}}=+1.
\label{eq:pecfresnel}
\end{equation}
For a finite-conductivity half-space, $r_c^p$ is obtained from
Eq.~\eqref{eq:fresnel} after replacing the semiconductor permittivity by
the reflector permittivity. The formulas below therefore apply to both
reflector models.

On the imaginary axis, $\omega=\ii\xi$ and $\kappa=\sqrt{k^2+\xi^2/c^2}$. Passivity and causality imply $\eps(\ii\xi)>0$ for the scalar dielectric model used here, so the equilibrium integral is nonsingular.

\subsection{Equilibrium pressure}

For a planar gap, the finite-temperature free energy per unit area is
\begin{equation}
\begin{aligned}
\mathcal F(a)={}&\frac{\kb T}{2\pi}\sum_{n=0}^{\infty}{}'
\int_0^\infty k\,\dd k\sum_{p}\\
&\times\ln\left[1-r_s^p(\ii\xi_n,k)r_c^p
e^{-2\kappa_n a}\right],
\end{aligned}
\label{eq:freeenergy}
\end{equation}
where $p\in\{\mathrm{TE},\mathrm{TM}\}$, $\xi_n=2\pi n\kb T/\hbar$, the prime gives the $n=0$ term half weight, and $\kappa_n=\sqrt{k^2+\xi_n^2/c^2}$. Differentiating with respect to $a$ gives the pressure on the reflector,
\begin{align}
\Peq(a)=&-\frac{\partial\mathcal F}{\partial a}
\label{eq:pwq}
\\
=&-\frac{\kb T}{\pi}\sum_{n=0}^{\infty}{}'
\int_0^\infty k\,\dd k\,\kappa_n
\sum_p\frac{R_{p,n}e^{-2\kappa_n a}}
{1-R_{p,n}e^{-2\kappa_n a}},
\label{eq:peq}
\end{align}
where $R_{p,n}=r_s^p(\ii\xi_n,k)r_c^p$. The differentiation is elementary because the single-gap reflection amplitudes are independent of $a$; this avoids the double counting that occurs if one differentiates a composite reflection operator that already contains the gap propagation factor. The differentiation is written out in the Supplemental Material \cite{SupplementalMaterial}.

For numerical quadrature we use
\begin{equation}
u=\kappa_n a,
\end{equation}
which converts Eq.~\eqref{eq:peq} into an exponentially convergent integral proportional to $a^{-3}$ at each Matsubara frequency.

\subsection{Driven occupation and physical bias margin}

The excess occupation above the unbiased state is modeled by the generalized Planck distribution \cite{Wurfel1982,ChenFan2016,Spreng2024},
\begin{align}
n(E,T,\eta)&=\frac{1}{\exp[(E-\eta E_g)/(\kb T)]-1},\qquad E\ge E_g,\\
\Delta n(E,T,\eta)&=n(E,T,\eta)-n(E,T,0).
\label{eq:occupation}
\end{align}
Here $E=\hbar\omega$; when numerical integrals use electron-volts, the conversion factor $e=1.602176634\times10^{-19}$ J/eV is inserted explicitly.

The occupation diverges as $\mu_\gamma\rightarrow E_g^-$. We therefore do not treat $\eta\rightarrow1$ as an optimization resource. Instead, the complete drive-uncertainty interval must satisfy
\begin{equation}
\eta+\Delta\eta\leq \eta_{\mathrm{phys}}=0.950.
\label{eq:margin}
\end{equation}
At 300 K and $E_g=1.424$ eV, this cap leaves
$E_g-\mu_\gamma=71.2$ meV $=2.75\kb T$ at the worst upper-drive endpoint. The number $2.75$ is therefore a consequence of the declared drive cap, not a separate fitted constant. The constitutive dielectric function itself remains passive; the drive modifies source occupation but does not insert optical gain into $\eps(\omega)$. An active dielectric model would require an additional pole-stability analysis.

The occupation is strictly increasing with $\eta$ below threshold. Indeed, with $x=(E-\eta E_g)/(\kb T)>0$,
\begin{equation}
\frac{\partial n}{\partial\eta}
=\frac{E_g}{\kb T}\frac{e^x}{(e^x-1)^2}>0.
\label{eq:monotonicn}
\end{equation}
The full pressure need not be globally monotonic because propagating and evanescent spectral weights can have different signs. We therefore do not assume monotonicity of $P$ a priori; Sec.~\ref{sec:optimization} establishes it over each final uncertainty interval by checking the resolved total spectral kernel.

\subsection{Propagating and evanescent nonequilibrium terms}

For a source body characterized by the excess occupation in Eq.~\eqref{eq:occupation}, the planar scattering formulation of nonequilibrium fluctuational electrodynamics gives \cite{MessinaAntezza2011,Kruger2012,ChenFan2016,Spreng2024}
\begin{widetext}
\begin{align}
\PPW(a,\eta)=&\frac{\hbar}{4\pi^2}\int_{\omega_g}^{\infty}\dd\omega\,\Delta n(\hbar\omega,T,\eta)
\int_0^{\omega/c} k\,\dd k\,k_z
\sum_p\frac{\left(1-|r_s^p|^2\right)\left(1+|r_c^p|^2\right)}
{\left|1-r_s^p r_c^p e^{2\ii k_z a}\right|^2},
\label{eq:ppw}\\
\PEW(a,\eta)=&-\frac{\hbar}{\pi^2}\int_{\omega_g}^{\infty}\dd\omega\,\Delta n(\hbar\omega,T,\eta)
\int_{\omega/c}^{\infty} k\,\dd k\,\operatorname{Im}k_z\,e^{-2a\operatorname{Im}k_z}
\sum_p\frac{\operatorname{Im}r_s^p\,\operatorname{Re}r_c^p}
{\left|1-r_s^p r_c^p e^{2\ii k_z a}\right|^2},
\label{eq:pew}
\end{align}
\end{widetext}
where $\omega_g=E_g/\hbar$. For a PEC, $|r_c^p|=1$, so the propagating contribution is nonnegative for a passive semiconductor and is the dominant repulsive term in the optimized regime. For any passive reflector the propagating numerator remains nonnegative. The evanescent contribution can have either sign.

Two changes of variables make the quadrature stable. In the propagating sector set
\begin{equation}
t=\frac{k_z}{k_0}\in[0,1],\qquad k_0=\frac{\omega}{c},
\end{equation}
so that
\begin{equation}
k\,\dd k\,k_z=k_0^3t^2\dd t.
\label{eq:pwtransform}
\end{equation}
In the evanescent sector set
\begin{equation}
\alpha=\frac{\operatorname{Im}k_z}{k_0}\in[0,\infty),
\end{equation}
which gives
\begin{equation}
k\,\dd k\,\operatorname{Im}k_z=k_0^3\alpha^2\dd\alpha.
\label{eq:ewtransform}
\end{equation}
These variables regularize the light-line endpoint and resolve the Fabry--P\'erot denominator without an ad hoc spectral-radius cap.

\subsection{Far-field plateau and cavity contrast}
\label{sec:farfield}

A pressure-only optimization over an unrestricted separation range has a conceptual defect: the attractive equilibrium and evanescent terms decay, while the propagating nonequilibrium pressure approaches a finite radiation-pressure plateau. Without an additional condition the optimizer can therefore prefer an arbitrarily large gap, which is not a cavity-design result. Bounds on the equilibrium force between passive mirrors are known \cite{LambrechtJaekelReynaud1997}; the result below instead concerns the driven, phase-averaged far-field contribution.

Let $\rho=r_s^p r_c^p$ with $|\rho|<1$. Phase averaging over many Fabry--P\'erot oscillations gives
\begin{equation}
\frac{1}{2\pi}\int_0^{2\pi}\frac{\dd\phi}{|1-\rho e^{\ii\phi}|^2}
=\frac{1}{1-|\rho|^2}.
\label{eq:poisson}
\end{equation}
For an arbitrary passive planar reflector the plateau is therefore
\begin{widetext}
\begin{equation}
P_{\infty}^{(c)}=
\frac{e}{4\pi^2}\int_{E_g}^{\infty}\dd E\,
\left(\frac{eE}{\hbar c}\right)^3\Delta n(E,T,\eta)
\int_0^1\dd t\,t^2\sum_p
\frac{(1-|r_s^p|^2)(1+|r_c^p|^2)}
{1-|r_s^p r_c^p|^2}.
\label{eq:pinfgeneral}
\end{equation}
\end{widetext}
This expression also gives a useful bound. Put
$x=|r_s^p|^2$ and $y=|r_c^p|^2$. For every propagating channel of passive bodies, $0\leq x,y\leq1$, and
\begin{equation}
F(x,y)=\frac{(1-x)(1+y)}{1-xy},\qquad
\frac{\partial F}{\partial y}=
\frac{1-x^2}{(1-xy)^2}\geq0.
\label{eq:reflectorbound}
\end{equation}
Consequently $1-x\leq F(x,y)\leq2$, with the upper value attained by a PEC. The PEC plateau is thus a modewise upper bound among passive specular reflectors for a fixed emitter spectrum. This statement concerns the phase-averaged far field; no analogous ordering is implied for the finite-gap force, whose cavity phase also changes with the reflector response.

For a PEC, $|r_c^p|=1$ and the factor $1-|r_s^p|^2$ cancels the phase-averaged denominator. Summing TE and TM gives a factor four, while $\int_0^1t^2\dd t=1/3$. Equation~\eqref{eq:pinfgeneral} then reduces to
\begin{equation}
\Pinf(E_g,\eta,T)=\frac{\hbar}{3\pi^2}\int_{\omega_g}^{\infty}\dd\omega\,
\left(\frac{\omega}{c}\right)^3\Delta n(\hbar\omega,T,\eta).
\label{eq:pinfomega}
\end{equation}
Using photon energy $E$ in electron-volts,
\begin{equation}
\Pinf=\frac{e}{3\pi^2}\int_{E_g}^{\infty}\dd E\,
\left(\frac{eE}{\hbar c}\right)^3\Delta n(E,T,\eta).
\label{eq:pinfev}
\end{equation}

We therefore define a robust cavity excess
\begin{equation}
C_{\mathrm{cav}}(\bm x;\mathcal R)=
J(\bm x;\mathcal R)-P_{\infty}^{(c)}(\mathcal R,E_g,\eta-\Delta\eta,T)
\label{eq:cavitycontrast}
\end{equation}
and require
\begin{equation}
C_{\mathrm{cav}}\ge0.10~\mathrm{mPa}.
\label{eq:cavityconstraint}
\end{equation}
This threshold is not a universal physical constant; it is a declared design specification chosen to force the optimum to retain a finite cavity-interference contribution. The same threshold is applied to the PEC and gold calculations, each relative to its own plateau.

\section{Finite central slab and controlled thick-slab reduction}
\label{sec:thickness}

Equations~\eqref{eq:peq}--\eqref{eq:pew} describe a semi-infinite emitting semiconductor. The central body in Fig.~\ref{fig:geometry} is finite, so a reduction criterion is needed before Eq.~\eqref{eq:bilateral} can be used independently on its two faces. Finite-thickness effects in nonequilibrium planar forces can otherwise remain important even when an infinite-half-space expression appears plausible \cite{RubioLopez2018}.

For propagating emission we define an emission-weighted single-pass leakage indicator
\begin{widetext}
\begin{equation}
\Lambda_{\mathrm{em}}(b)=
\frac{\displaystyle\int_{E_g}^{\infty}\dd E\,\Delta n\,k_0^3
\int_0^1\dd q\,q\sqrt{1-q^2}\sum_p(1-|r_s^p|^2)
\exp[-2k_0\operatorname{Im}(s_z)b]}
{\displaystyle\int_{E_g}^{\infty}\dd E\,\Delta n\,k_0^3
\int_0^1\dd q\,q\sqrt{1-q^2}\sum_p(1-|r_s^p|^2)},
\qquad s_z=\sqrt{\eps(E)-q^2},
\label{eq:leakage}
\end{equation}
\end{widetext}
with $q=k/k_0$. We require $\Lambda_{\mathrm{em}}\le10^{-3}$ at the largest allowed drive $\eta=0.95$.

A separate conservative indicator is used for equilibrium paths that traverse the slab and couple the two vacuum gaps. Ignoring reflection losses can only overestimate such a path. With the dimensionless equilibrium variable $u=\kappa a$, adding an extra decay length $b$ yields the ratio
\begin{equation}
C_{\mathrm{eq}}(b;a)=
\frac{\int_0^\infty u^2e^{-2u}e^{-2(b/a)u}\dd u}
{\int_0^\infty u^2e^{-2u}\dd u}
=\left(1+\frac{b}{a}\right)^{-3},
\label{eq:eqcoupling}
\end{equation}
where the last equality follows from $\int_0^\infty u^2e^{-\beta u}\dd u=2/\beta^3$. Requiring $C_{\mathrm{eq}}\le10^{-3}$ at the largest design gap $a_{\max}=1.50~\mu\mathrm m$ gives
\begin{equation}
b\ge9a_{\max}=13.5~\mu\mathrm m.
\label{eq:bmin}
\end{equation}

For GaAs the emission criterion alone reaches $10^{-3}$ at approximately $b=10.45~\mu\mathrm{m}$, whereas the conservative equilibrium indicator requires $b\geq 9a_{\max}=13.5~\mu\mathrm{m}$. We therefore adopt $b=13.5~\mu\mathrm{m}$, for which $\Lambda_{\mathrm{em}}=1.77\times10^{-4}$ and $C_{\mathrm{eq}}=10^{-3}$. At $b=100~\mathrm{nm}$ the emission-weighted leakage is approximately $0.92$, so a 100-nm central layer does not lie in the opaque-emitter regime assumed in Eqs.~\eqref{eq:ppw} and \eqref{eq:pew}. The adopted thickness is selected lexicographically as the smallest value satisfying the stated reduction criteria; it is not obtained by maximizing a pressure that remains explicitly dependent on $b$. Once these criteria are satisfied, the reduced pressure model is independent of $b$. The indicators are prospective reduction tests rather than exact errors. Their thickness dependence and threshold study are given in the Supplemental Material \cite{SupplementalMaterial}; the full calculation below supplies the a posteriori check.

\subsection{Full finite-slab validation}
\label{sec:finitevalidation}

To verify the independent-face reduction directly, we also evaluate the complete finite-thickness scattering problem at the nominal optimum and at the active robust uncertainty point. The calculation retains the reflection and transmission amplitudes of the finite GaAs slab, the correlations between the fields emitted from its two faces, and all multiple reflections involving both PEC--slab gaps. The corresponding two-port scattering formulation is given in the Supplemental Material \cite{SupplementalMaterial}.

At the nominal point we set $a_L=a_R=a^\star$ and $\eta=0.945$. At the lower-drive uncertainty endpoint, the three left--right-symmetry-distinct gap combinations $(a^\star-\Delta a,a^\star-\Delta a)$,
$(a^\star-\Delta a,a^\star+\Delta a)$, and $(a^\star+\Delta a,a^\star+\Delta a)$ were evaluated. The mixed-gap configuration gives the smallest full finite-slab bilateral pressure and is
reported in Table~\ref{tab:finiteslab}. Here $\Delta P_{\min}=P_{\min}^{\mathrm{fs}}-P_{\min}^{\mathrm{red}}$.

\begin{widetext}
\begin{center}
\begin{table}[t]
\caption{Comparison of the independent-face reduced model with the full finite-slab calculation for the PEC reference problem. All pressures are in mPa. At the nominal point symmetry gives $P_L^{\mathrm{fs}}=P_R^{\mathrm{fs}}$. The two orientations of the mixed-gap worst-case configuration are equivalent under interchange of the reflector labels. Extra digits are retained only to resolve the small finite-thickness correction.}
\label{tab:finiteslab}
\begin{ruledtabular}
\begin{tabular}{lcccccc}
Point & $(a_L,a_R)$ ($\mu$m) & $\eta$ &
$P_{\min}^{\mathrm{red}}$ &
$P_L^{\mathrm{fs}}$ &
$P_R^{\mathrm{fs}}$ &
$\Delta P_{\min}$\\
\hline
Nominal & $(1.351,1.351)$ & 0.945 &
3.056574 & 3.056547 & 3.056547 &
$-2.68\times10^{-5}$\\
Active worst case & $(1.341,1.361)$ & 0.940 &
2.256837 & 2.256812 & 2.256801 &
$-3.62\times10^{-5}$\\
\end{tabular}
\end{ruledtabular}
\end{table}
\end{center}
\end{widetext}

The maximum absolute discrepancy is $3.62\times10^{-5}~\mathrm{mPa}$, corresponding to $1.60\times10^{-3}\%$ of the reduced-model pressure. It is far below both the displayed precision of the headline pressures and the adopted $10^{-3}$ reduction-indicator threshold. The full calculation therefore confirms that cross-slab transmission and correlated two-face emission are negligible at the reported design, while leaving the reduced model useful for the global optimization.

\section{Robust constrained optimization}
\label{sec:optimization}

\subsection{Design variables and uncertainty set}

The conceptual design vector is
\begin{equation}
\bm x=(a_L,a_R,b,\eta,E_g).
\end{equation}
The continuous band gap is first optimized using the analytically derived far-field scale in Eq.~\eqref{eq:pinfev}; a real material is then selected and its complete optical response is used in the finite-gap calculation. For the GaAs stage, $E_g=1.424$ eV is fixed and the remaining design variables are $(a_L,a_R,b,\eta)$.

We impose the compact design box
\begin{align}
0.25~\mu\mathrm m&\le a_L,a_R\le1.50~\mu\mathrm m,\label{eq:gapbounds}\\
0.85&\le\eta\le0.945.\label{eq:etabounds}
\end{align}
The upper gap is a microcavity design limit rather than an asymptotic cutoff; Eq.~\eqref{eq:cavityconstraint} independently rejects designs whose performance is explained almost entirely by the far-field plateau. The drive bound ensures that the uncertainty interval remains below $\eta=0.95$.

Fabrication and drive uncertainty are represented by
\begin{equation}
\begin{aligned}
\mathcal U=\bigl\{&(\delta a_L,\delta a_R,\delta\eta):
|\delta a_L|,|\delta a_R|\le10~\mathrm{nm},\\
&|\delta\eta|\le0.005\bigr\}.
\end{aligned}
\label{eq:uncertainty}
\end{equation}
The robust bilateral objective is
\begin{equation}
\begin{aligned}
J(\bm x;\mathcal R)=\min_{\bm\delta\in\mathcal U}
\min\!\bigl\{&P(a_L+\delta a_L,\eta+\delta\eta),\\
&P(a_R+\delta a_R,\eta+\delta\eta)\bigr\}.
\end{aligned}
\label{eq:robustobj}
\end{equation}
Thus a candidate fails if either reflector becomes attractive anywhere in the uncertainty box. The reflector argument has been suppressed inside the braces for readability.

The uncertainty set in Eq.~\eqref{eq:uncertainty} is continuous. We do not identify a dense tensor grid with a proof over that set. Write the nonequilibrium pressure as
\begin{equation}
\begin{aligned}
P(a,\eta)={}&P_{\rm eq}(a)+
\int_{E_g}^{\infty}K(E,a)\Delta n(E,\eta)\,\dd E,\\
K={}&K_{\rm PW}+K_{\rm EW}.
\end{aligned}
\label{eq:spectralkernel}
\end{equation}
Because $\partial_\eta\Delta n>0$, nonnegativity of $K(E,a)$ throughout an admitted gap interval is sufficient for $P$ to increase with drive. For both optimized reflector models, the resolved total kernel is positive throughout the energy window and gap-uncertainty interval; the smallest calculated $\partial P/\partial\eta$ is $136.7~\mathrm{mPa}$ for the PEC and $131.9~\mathrm{mPa}$ for gold per unit change in $\eta$. The inner drive minimum is therefore at $\eta-\Delta\eta$. At that drive endpoint we locate every sign change of the sampled gap derivative in the closed interval, refine each candidate by local quadratic interpolation, and compare all candidates with both interval endpoints. The only interior stationary candidate is a local maximum, and the two endpoints are equal at the refined robust design. For the PEC reference calculation, nested tensor grids up to $81\times41$ points are retained as an independent diagnostic. The detailed audit is supplied in the Supplemental Material \cite{SupplementalMaterial}.

The constrained problem is
\begin{align}
&\underset{a_L,a_R,b,\eta}{\text{maximize}}&&J(\bm x),\label{eq:optmain}\\
&\text{subject to}&&\text{Eqs.~\eqref{eq:margin}, \eqref{eq:cavityconstraint}, \eqref{eq:gapbounds}, \eqref{eq:etabounds}},\nonumber\\
&&&\Lambda_{\mathrm{em}}(b)\le10^{-3},\quad
C_{\mathrm{eq}}(b;a_{\max})\le10^{-3}.\nonumber
\end{align}
After the maximum pressure is found, a lexicographic second stage minimizes $b$ without degrading the optimal value of $J$ within numerical tolerance. This prevents unnecessary thickness from being rewarded once the two faces are effectively decoupled.

\subsection{Existence of an optimum}

The formulation deliberately avoids a formal optimization over an open or singular domain. The gap and drive intervals are closed and bounded. The margin in Eq.~\eqref{eq:margin} keeps the Bose factor finite, and the passive optical model keeps all scattering denominators finite away from integrable cavity resonances. The quadrature integrands are continuous in $(a,\eta)$ and dominated by exponentially or thermally decaying envelopes. Therefore $P(a,\eta)$ is continuous on the feasible design box. The minimum over the compact uncertainty set in Eq.~\eqref{eq:robustobj} preserves continuity. The thickness and cavity-contrast inequalities define closed subsets of the compact box (with $b$ restricted to any finite fabrication interval containing the active constraint). The feasible set is therefore compact, and by the Weierstrass theorem $J$ attains a maximum whenever the feasible set is nonempty.

\subsection{Symmetry theorem for the bilateral optimum}
\label{sec:symmetry}

The numerical search need not assume a centered slab. Symmetry follows from the objective after the thick-slab reduction.

Define the one-gap robust value at fixed $\eta$ and reflector $\mathcal R$,
\begin{equation}
j(a,\eta;\mathcal R)=\min_{|\delta a|\le\Delta a,\,|\delta\eta|\le\Delta\eta}
P(a+\delta a,\eta+\delta\eta;\mathcal R).
\label{eq:jone}
\end{equation}
Because the left and right reflectors are identical and the uncertainty bounds are the same,
\begin{equation}
J(a_L,a_R,\eta;\mathcal R)=
\min\{j(a_L,\eta;\mathcal R),j(a_R,\eta;\mathcal R)\}.
\label{eq:Jmin}
\end{equation}
Let $j_\star(\eta;\mathcal R)=\max_a j(a,\eta;\mathcal R)$. For any $(a_L,a_R)$,
\begin{equation}
J(a_L,a_R,\eta;\mathcal R)\le j_\star(\eta;\mathcal R).
\end{equation}
Choose any maximizer $a_\star$ of $j(a,\eta;\mathcal R)$ and set $a_L=a_R=a_\star$. Then
\begin{equation}
J(a_\star,a_\star,\eta;\mathcal R)=j_\star(\eta;\mathcal R),
\end{equation}
which saturates the bound. Hence at least one global bilateral optimizer is symmetric. This result does not assume that an off-center geometry is unphysical; it proves that asymmetry cannot improve the max--min objective for identical reflectors and identical uncertainty sets once the two slab faces are decoupled.

\subsection{Small-uncertainty structure}
\label{sec:smalluncertainty}

The different roles of gap and drive uncertainty can be obtained without
solving the complete numerical problem. Suppose the positivity audit of
Eq.~\eqref{eq:spectralkernel} places the active drive at
$\eta_-=\eta_0-\Delta\eta$, and define
$f(a)=P(a,\eta_-;\mathcal R)$. Let $a_m$ be a nondegenerate local maximum
of the selected repulsive lobe, so that $f'(a_m)=0$ and $f''(a_m)<0$.
Write the robust center as $a_0=a_m+\epsilon$. If the admitted interval
contains no local minimum, its worst value is at one of its endpoints and
the max--min center equalizes them. Taylor expansion gives
\begin{align}
0={}&f(a_m+\epsilon+\Delta a)-f(a_m+\epsilon-\Delta a)\nonumber\\
={}&2f''(a_m)\epsilon\Delta a+
\frac{1}{3}f'''(a_m)(\Delta a)^3+O[(\Delta a)^5].
\end{align}
Consequently,
\begin{align}
a_0-a_m&=-\frac{f'''(a_m)}{6f''(a_m)}(\Delta a)^2
+O[(\Delta a)^4],
\label{eq:robustshift}\\
J_{\rm gap}&=f(a_m)+\frac{1}{2}f''(a_m)(\Delta a)^2
+O[(\Delta a)^4].
\label{eq:gappenalty}
\end{align}
Symmetric gap uncertainty therefore shifts the lobe center and lowers the
optimized pressure only at second order. In contrast, the envelope theorem
and Eq.~\eqref{eq:monotonicn} give
\begin{equation}
J(\Delta\eta)=J(0)-
\left.\frac{\partial P}{\partial\eta}\right|_{a_m,\eta_0}
\Delta\eta+O[(\Delta\eta)^2],
\label{eq:drivepenalty}
\end{equation}
so drive uncertainty enters at first order. For the PEC lobe at
$\eta_-=0.940$, Eqs.~\eqref{eq:robustshift} and \eqref{eq:gappenalty}
predict a $0.550~\mathrm{nm}$ shift and a $0.00509~\mathrm{mPa}$ gap
penalty. Direct reoptimization gives $0.548~\mathrm{nm}$ and
$0.00506~\mathrm{mPa}$, respectively. This asymptotic result explains the
uncertainty hierarchy reported below and is not specific to the adopted
GaAs response.

\section{Material response and numerical method}
\label{sec:numerics}

\subsection{Band-gap preoptimization and GaAs selection}

Before invoking a detailed optical model, Eq.~\eqref{eq:pinfev} provides a material-independent screen for the band-gap scale that maximizes the asymptotic momentum flux of an ideal opaque direct-gap emitter. The continuous optimum is $E_g=1.229$ eV at the robust low-drive endpoint $\eta=0.94$ and $1.344$ eV at the nominal drive $\eta=0.945$. GaAs, with $E_g=1.424$ eV, lies only $0.46\%$ below the ideal pressure optimum at $\eta=0.945$ and has both a well-established analytic response and a direct benchmark in the biased-Casimir literature \cite{Spreng2024,Adachi1989,Polyanskiy2024}. We therefore use GaAs for the finite-gap calculation. The full material screen is reported in the Supplemental Material \cite{SupplementalMaterial}.

The screen also explains why an InSb baseline is poorly matched to the present 300-K, submicron-to-low-micron biased-emission objective: its small band gap produces a much smaller momentum-flux scale at the same fractional chemical potential. This conclusion concerns the driven interband mechanism studied here and does not imply that InSb is generally inferior for magneto-optical or nonreciprocal Casimir physics.

\subsection{Causal GaAs dielectric model}

At real frequency we use the analytic Adachi parameterization for GaAs \cite{Adachi1989}, with optical-constant cross-checks against the refractiveindex.info compilation \cite{Polyanskiy2024}. The numerical parameters used in the code are $E_0=1.42$ eV, $E_1=2.90$ eV, $E_2=4.7$ eV, and the corresponding broadenings and critical-point amplitudes of the Adachi model.

The equilibrium calculation requires $\eps(\ii\xi)$ rather than a direct substitution of imaginary frequency into a piecewise real-axis fit. We therefore construct it causally from the absorptive spectrum,
\begin{widetext}
\begin{equation}
\eps(\ii\xi)=1+\frac{2}{\pi}\int_0^{E_c}\dd E\,
\frac{E\eps_2(E)}{E^2+X^2}
+\Delta\eps_{\mathrm{UV}}\frac{E_{\mathrm{UV}}^2}{E_{\mathrm{UV}}^2+X^2}
+\Delta\eps_{\mathrm{ph}}\frac{E_{\mathrm{ph}}^2}{E_{\mathrm{ph}}^2+X^2},
\label{eq:kk}
\end{equation}
\end{widetext}
where $X=\hbar\xi$ expressed in eV and $E_c=6$ eV. The first term is the Kramers--Kronig continuation of the Adachi absorptive spectrum. A high-energy oscillator at $E_{\mathrm{UV}}=10$ eV restores $\eps_\infty=10.89$, and a far-infrared oscillator at $E_{\mathrm{ph}}=0.035$ eV restores $\eps(0)=12.9$. For the adopted numerical spectrum the truncated interband integral gives $9.600$, so $\Delta\eps_{\mathrm{UV}}=1.290$ and $\Delta\eps_{\mathrm{ph}}=2.01$. This completion preserves causality and the required low-frequency limits without extrapolating the Adachi critical-point formula outside its intended spectral window.

Optical-model dependence is checked with the Raki\'{c}--Majewski extension of the GaAs critical-point model \cite{RakicMajewski1996}. Its published real-frequency parameterization is used directly, and its imaginary-axis response is reconstructed by the same Kramers--Kronig procedure, now integrating the absorptive spectrum to $8$ eV and restoring the same static and high-frequency limits. This is a second published causal parameterization rather than a uniform rescaling of the primary susceptibility. Model formulas and parameters are collected in the Supplemental Material \cite{SupplementalMaterial}.

\subsection{Finite-conductivity gold benchmark}
\label{sec:goldmodel}

The nonideal reflector is semi-infinite gold described by the causal Lorentz--Drude fit of Raki\'{c} \emph{et al.} \cite{Rakic1998},
\begin{equation}
\eps_{\rm Au}(E)=1-
\frac{f_0E_p^2}{E(E+\ii\Gamma_0)}+
\sum_{j=1}^{5}\frac{f_jE_p^2}
{E_j^2-E^2-\ii\Gamma_jE}.
\label{eq:goldreal}
\end{equation}
All energies are in electron-volts. We use $E_p=9.03$, $f_0=0.760$, and $\Gamma_0=0.053$, together with
\begin{align}
(f_1,\ldots,f_5)&=(0.024,0.010,0.071,0.601,4.384),\nonumber\\
(E_1,\ldots,E_5)&=(0.415,0.830,2.969,4.304,13.32),\nonumber\\
(\Gamma_1,\ldots,\Gamma_5)&=(0.241,0.345,0.870,2.494,2.214).
\label{eq:goldparameters}
\end{align}
The same causal model is continued to the imaginary axis as
\begin{equation}
\eps_{\rm Au}(\ii X)=1+
\frac{f_0E_p^2}{X(X+\Gamma_0)}+
\sum_{j=1}^{5}\frac{f_jE_p^2}
{E_j^2+X^2+\Gamma_jX}.
\label{eq:goldimag}
\end{equation}
The zero-frequency TE term follows the dissipative Drude prescription and vanishes; the TM reflection amplitude tends to unity. This model is used to test survival of the optimized mechanism under finite reflectivity and absorption. It is not intended to remove sample-to-sample uncertainty in the optical response of a fabricated gold film.

\subsection{Search strategy and numerical verification}

The propagating pressure contains Fabry--P\'erot oscillations, so a single
local optimization from an arbitrary starting point is not reliable. We use
a staged deterministic search. First, a rectangular grid spanning the
complete declared box
$0.25\leq a_0\leq1.50~\mu\mathrm{m}$ and
$0.85\leq\eta_0\leq0.945$ identifies all repulsive lobes that can satisfy
the cavity-contrast constraint. The four-corner uncertainty value is used
only as a screening bound at this stage. Second, the best surviving lobe
is refined by balancing the two gap-error endpoints at the lower drive
endpoint,
\begin{equation}
P(a_\star-\Delta a,\eta_\star-\Delta\eta)
=
P(a_\star+\Delta a,\eta_\star-\Delta\eta).
\label{eq:endpointbalance}
\end{equation}
Third, the drive-monotonicity and stationary-candidate audit described in
Sec.~\ref{sec:optimization} resolves the inner minimization, with the nested
grids used as an independent check. Finally, the symmetry theorem sets
$a_L=a_R=a_\star$, and the slab thickness is chosen by the lexicographic
reduction criterion of Sec.~\ref{sec:thickness}.

The full two-dimensional coarse map is retained for the PEC reference and
for Fig.~\ref{fig:map}. Once the positive drive derivative has been verified,
the gold reoptimization needs only a complete one-dimensional scan of the
declared gap interval at the upper nominal drive, followed by the same
endpoint-balancing refinement. No PEC optimum is used as the gold starting
point. The alternative GaAs model is evaluated independently at the nominal
and active worst-case PEC points. A secondary $\pm5\%$ causal susceptibility
rescaling is retained in the machine-readable data but is not used as the
principal optical-model validation.

The photon-energy integral is evaluated from $E_g$ to
$E_g+0.60~\mathrm{eV}$, and the evanescent variable is truncated at
$\alpha_{\max}=10$. Numerical verification is performed in two parts.
First, the quadrature orders $N_u$, $N_E$, $N_{\rm PW}$, and $N_{\rm EW}$
are varied one at a time while the remaining orders are held at their final
values. Second, the photon-energy window and $\alpha_{\max}$ are varied
independently to test truncation convergence. The two highest settings in
every one-factor test agree to better than $4\times10^{-10}$ in relative
total pressure. The complete sequences and a compact convergence table are
supplied in the Supplemental Material and as machine-readable data
\cite{SupplementalMaterial}.

\section{Optimization results}
\label{sec:results}

\subsection{Pressure curves and feasible repulsive lobes}

\begin{figure}[t]
\includegraphics[width=0.9\columnwidth]{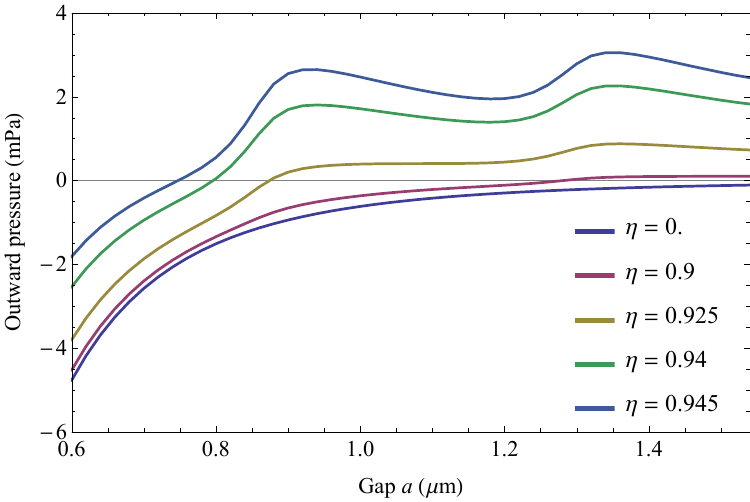}
\caption{Outward pressure on one PEC versus vacuum gap for several drive values $\eta=\mu_\gamma/E_g$. The passive curve is attractive; forward bias introduces oscillatory repulsive windows through propagating-wave Fabry--P\'erot interference.}
\label{fig:curves}
\end{figure}

Figure~\ref{fig:curves} shows the total one-gap pressure as the drive is increased. At zero bias only the equilibrium contribution remains and the force is attractive. Bias generates a positive propagating term, producing oscillatory sign reversals associated with the cavity phase. The magnitude and location of these lobes make it inappropriate to optimize by a single local derivative from an arbitrary starting point.

\begin{figure}[t]
\includegraphics[width=0.9\columnwidth]{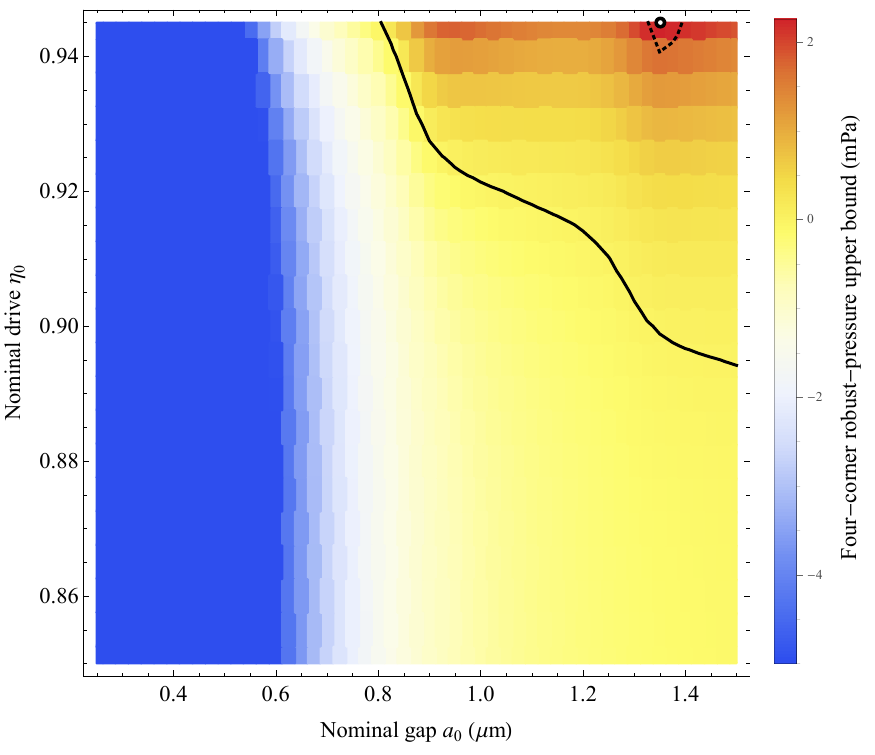}
\caption{Four-corner screening bound for the PEC robust bilateral pressure over the complete $(a_0,\eta_0)$ design box. Because the full uncertainty set contains its corners, this quantity is an upper bound on the continuous-uncertainty objective and is used to locate and exclude candidate lobes efficiently. The marker denotes the refined design, whose objective is subsequently checked by the drive-monotonicity and stationary-candidate audit described in the text. Solid and dotted contours denote zero screening pressure and the imposed $0.10~\mathrm{mPa}$ cavity-excess threshold, respectively. Pressure values below $-5~\mathrm{mPa}$ are saturated at $-5~\mathrm{mPa}$ in the color scale to preserve contrast near the
feasible region.}
\label{fig:map}
\end{figure}

The robust design map is shown in Fig.~\ref{fig:map}. Each pixel is the worst one-gap pressure over $\delta a=\pm10$ nm and $\delta\eta=\pm0.005$; by the theorem of Sec.~\ref{sec:symmetry}, this is also the optimal bilateral value when both gaps are set to the same nominal point. The zero-pressure contour separates robust attraction from repulsion. The additional cavity-contrast contour prevents selection of a region whose pressure is indistinguishable, at the design level, from the asymptotic plateau.

The optimum is
\begin{equation}
 a_L^\star=a_R^\star=1.351~\mu\mathrm m,
 \quad \eta^\star=0.945,
 \quad b^\star=13.5~\mu\mathrm m.
\label{eq:optimum}
\end{equation}
The worst-case bilateral pressure is
\begin{equation}
J^\star=2.257~\mathrm{mPa}.
\label{eq:Jopt}
\end{equation}
The active pressure uncertainty occurs at the lower drive $\eta=0.940$ and one of the two gap endpoints. At the refined nominal gap, the two lower-drive endpoint pressures differ by only $3.2\times10^{-6}~\mathrm{mPa}$ in the exported continuous-uncertainty audit, well below the displayed precision of $J^\star$. This near equality is the numerical realization of Eq.~\eqref{eq:endpointbalance}.

\begin{figure*}[t]
\includegraphics[width=0.45\textwidth]{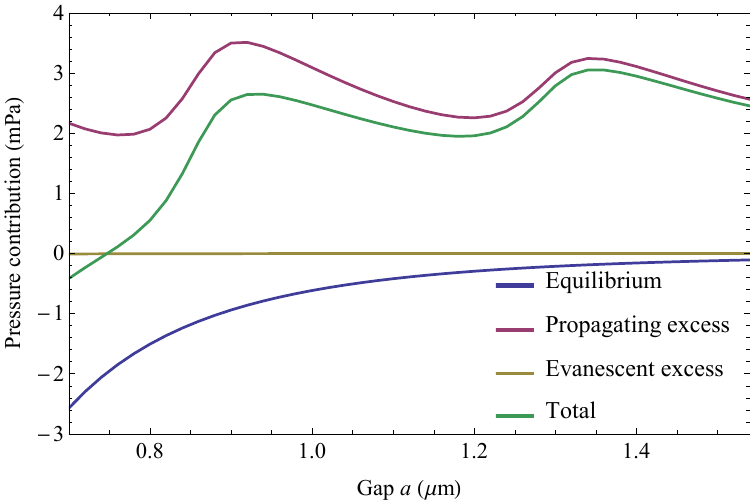}\hfill
\includegraphics[width=0.45\textwidth]{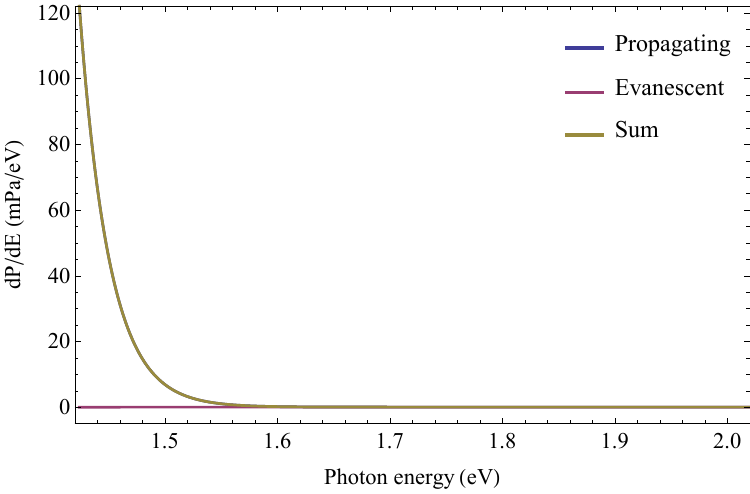}
\caption{Origin of the nonequilibrium repulsion at the nominal PEC optimum. Left: Pressure decomposition versus gap at $\eta=0.945$. The propagating term dominates, the equilibrium term remains attractive, and the evanescent correction is small. Right: Photon-energy-resolved nonequilibrium pressure. The driven contribution is concentrated near the direct band edge. The integrated evanescent term is $-1.210\times10^{-3}~\mathrm{mPa}$ and is visually unresolved on both plotted scales.}
\label{fig:mechanism}
\end{figure*}

At the nominal point the decomposition is
\begin{align}
\Peq&=-0.1842~\mathrm{mPa},\\
\PPW&=+3.242~\mathrm{mPa},\\
\PEW&=-1.210\times 10^{-3}~\mathrm{mPa},\\
\Ptot&=+3.057~\mathrm{mPa}.
\label{eq:decompnumbers}
\end{align}
Thus the repulsion is not a subtle cancellation of large evanescent terms. It is dominated by propagating photons emitted by the biased semiconductor and filtered by the PEC--semiconductor cavity. Figure~\ref{fig:mechanism} shows both the separation dependence and the photon-energy-resolved spectrum.

\subsection{Cross-evaluation and finite conductivity}
\label{sec:goldresults}

Table~\ref{tab:crosscomparison} cross-evaluates each nominal and robust
design. The nominal PEC optimum remains repulsive throughout the admitted
uncertainty box, but its certified worst-case pressure is
$2.255~\mathrm{mPa}$. Robust reoptimization moves the gap by only
$1.8~\mathrm{nm}$ and raises that value by $0.0023~\mathrm{mPa}$. The gold
calculation gives the same qualitative result: a $1.9~\mathrm{nm}$ shift
raises the worst-case pressure by $0.0023~\mathrm{mPa}$. Thus the present
max--min calculation does not manufacture repulsion or select another
Fabry--P\'erot lobe. Its principal role is to certify the bilateral pressure,
identify the active perturbation, and quantify the small improvement that is
available within the selected lobe.

\begin{table*}[t]
\caption{Apples-to-apples cross-evaluation of nominally and robustly optimized designs. $P_{\rm nom}$ is evaluated without perturbations; $J_{\rm worst}$ and $C_{\rm cav,worst}$ use $\Delta a=10~\mathrm{nm}$ and $\Delta\eta=0.005$. Gaps are in $\mu$m and pressures are in mPa.}
\label{tab:crosscomparison}
\begin{ruledtabular}
\begin{tabular}{llccccc}
Reflector & Design criterion & $a$ & $P_{\rm nom}$ & $J_{\rm worst}$ &
$C_{\rm cav,worst}$ & Both reflectors repulsive?\\
\hline
PEC & Nominal & 1.3494 & 3.0568 & 2.2546 & 0.1789 & Yes\\
PEC & Robust  & 1.3512 & 3.0566 & 2.2568 & 0.1812 & Yes\\
Au  & Nominal & 1.3232 & 2.9363 & 2.1620 & 0.1577 & Yes\\
Au  & Robust  & 1.3251 & 2.9361 & 2.1643 & 0.1600 & Yes\\
\end{tabular}
\end{ruledtabular}
\end{table*}

Finite conductivity shifts the robust gap by $26~\mathrm{nm}$ relative to
the PEC result and lowers the robust pressure by $4.1\%$, but it removes
neither repulsion nor the required cavity excess. The gold optimum retains
$95.9\%$ of the PEC robust pressure, and its reflector-specific plateau
obeys the passive-reflector bound in Eq.~\eqref{eq:reflectorbound}.

\subsection{Robustness trade-off and optical-model dependence}

The baseline result is not tied to one unexamined tolerance choice.
Table~\ref{tab:tradeoff} varies one specification at a time and reoptimizes
the PEC gap over the full declared box. Doubling the gap uncertainty from
$10$ to $20~\mathrm{nm}$ reduces the certified pressure by only
$0.0147~\mathrm{mPa}$, whereas halving the drive uncertainty recovers
$0.368~\mathrm{mPa}$. The design is therefore drive limited. The cavity
excess constraint is inactive between $0.05$ and $0.15~\mathrm{mPa}$; a
$0.20~\mathrm{mPa}$ requirement is infeasible in the declared box.

\begin{table}[t]
\caption{PEC robustness trade-off at $\eta_0=0.945$. Each row is independently reoptimized. Pressures are in mPa. A dash denotes no feasible design in the declared box.}
\label{tab:tradeoff}
\begin{ruledtabular}
\begin{tabular}{cccccc}
$\Delta a$ (nm) & $\Delta\eta$ & $C_{\rm req}$ & $a^\star$ ($\mu$m) & $J^\star$ & $C_{\rm cav}^\star$\\
\hline
5  & 0.0050 & 0.10 & 1.3508 & 2.2606 & 0.1849\\
10 & 0.0050 & 0.10 & 1.3512 & 2.2568 & 0.1812\\
20 & 0.0050 & 0.10 & 1.3528 & 2.2421 & 0.1664\\
10 & 0      & 0.10 & 1.3500 & 3.0498 & 0.2999\\
10 & 0.0025 & 0.10 & 1.3505 & 2.6249 & 0.2363\\
10 & 0.0050 & 0.05 & 1.3512 & 2.2568 & 0.1812\\
10 & 0.0050 & 0.15 & 1.3512 & 2.2568 & 0.1812\\
10 & 0.0050 & 0.20 & -- & -- & --\\
\end{tabular}
\end{ruledtabular}
\end{table}

Table~\ref{tab:gaasmodels} repeats the total-pressure calculation with the
Raki\'{c}--Majewski GaAs response while holding the reported geometry and
drive fixed. The largest total-pressure change is $0.0444\%$ at the nominal
point; at the active worst-case point it is $0.0329\%$. This agreement does
not replace sample-specific optical data, but it shows that the repulsive
sign and the reported margins are not peculiar to one analytic GaAs fit.

\begin{table}[t]
\caption{Optical-model comparison at the PEC robust design. The total pressure is in mPa; the percentage is relative to the Adachi value at the same point. Component-level values are supplied as machine-readable data. Percentage changes are calculated from the unrounded pressures.}
\label{tab:gaasmodels}
\begin{ruledtabular}
\begin{tabular}{llcc}
Point & GaAs response & $P_{\rm tot}$ & Change (\%)\\
\hline
Nominal & Adachi & 3.057 & 0\\
Nominal & Raki\'{c}--Majewski & 3.058 & 0.0444\\
Active worst case & Adachi & 2.257 & 0\\
Active worst case & Raki\'{c}--Majewski & 2.258 & 0.0329\\
\end{tabular}
\end{ruledtabular}
\end{table}

\subsection{How much of the optimum is genuinely cavity dependent?}

At $\eta=0.945$, Eq.~\eqref{eq:pinfev} gives the GaAs far-field plateau
\begin{equation}
\Pinf=2.750~\mathrm{mPa}.
\end{equation}
The nominal cavity excess is therefore
\begin{equation}
P(a^\star,0.945)-\Pinf=0.3067~\mathrm{mPa},
\end{equation}
or $11.2\%$ of the plateau. More importantly, at the worst drive $\eta=0.940$,
\begin{equation}
\Pinf=2.076~\mathrm{mPa},\qquad
C_{\mathrm{cav}}^\star=0.1812~\mathrm{mPa},
\end{equation}
so the robust optimum satisfies the $0.10$ mPa cavity-contrast requirement with margin. For gold the corresponding plateau and excess are $2.004$ and $0.160~\mathrm{mPa}$, respectively. The result should therefore be interpreted as a cavity-enhanced nonequilibrium fluctuation pressure with a substantial radiation-pressure component, not as a purely zero-point Casimir force.

\subsection{Spectral origin and bilateral symmetry}

Figure~\ref{fig:mechanism}(b) resolves the nonequilibrium pressure by photon energy. The excess occupation is concentrated just above the GaAs band edge, and the propagating spectrum carries essentially all of the positive pressure. This narrow spectral origin is physically distinct from replacing the biased semiconductor by a simple effective-temperature model, which would modify a broad thermal spectrum. A separate offset scan, shown in the Supplemental Material \cite{SupplementalMaterial}, confirms the bilateral consequence of the symmetry theorem: moving the slab at fixed mean gap raises one reflector pressure and lowers the other, so their minimum is maximized at zero offset.

\section{Physical interpretation and model scope}
\label{sec:interpretation}

Three distinctions are important for interpreting Eqs.~\eqref{eq:optimum} and \eqref{eq:Jopt}.

First, the calculation predicts a \emph{driven} nonequilibrium fluctuation pressure. The external bias supplies free energy, and the dominant positive term is carried by propagating photons. Calling the entire positive pressure a zero-temperature vacuum Casimir force would be misleading. The appropriate description is a nonequilibrium Casimir--Lifshitz or fluctuation-induced pressure in a cavity, consistent with the scattering formulations of Refs.~\cite{MessinaAntezza2011,Kruger2012,ChenFan2016,Spreng2024}.

Second, positive pressure on fixed boundaries is not by itself mechanical levitation. The reflectors must be position controlled or attached to mechanical supports, whose stiffness and force gradient determine dynamical stability. The present optimization guarantees robust sign and magnitude of the electromagnetic load; it does not claim a free-body stable equilibrium of the complete three-body assembly \cite{RahiKardarEmig2010,Kruger2011}.

Third, the PEC has two distinct roles. It makes the reference cavity analytically transparent and, by Eq.~\eqref{eq:reflectorbound}, gives a rigorous upper bound on the phase-averaged far-field pressure of any passive specular reflector. It is \emph{not} a monotonic bound on the finite-gap pressure, because finite conductivity changes the equilibrium term, momentum reversal, cavity linewidth, and cavity phase at the same time. The gold reoptimization is therefore an independent physical check rather than an empirical correction factor applied to the PEC answer.

The spatially uniform $\eta$ remains an ideal reservoir assumption. The calculation does not solve carrier injection, diffusion, recombination, or heating inside the GaAs slab. Those processes determine whether a given device realizes the assumed quasi-Fermi-level splitting, but they do not alter the mathematical distinction between nominal and robust electromagnetic pressures established here.

\section{Discussion}

The main contribution of this work is a bilateral robust formulation of nonequilibrium Casimir repulsion together with a nonideal-reflector test. The forces on the two reflectors are treated separately, and a design fails if either outward pressure falls below the objective under the admitted perturbations. In the controlled thick-slab limit, identical boundaries and uncertainty sets imply that the bilateral problem admits a symmetric optimizer. Equality of $a_L$ and $a_R$ is therefore a consequence of the reduced max--min problem. It is not a hidden centering assumption. This conclusion need not survive unequal reflectors, asymmetric drive, or appreciable transmission through the central slab, for which the full two-gap scattering problem must be restored.

The optimum is conditional on the stated feasible set. In particular, $\eta=0.945$ lies at the upper nominal-drive bound and should not be interpreted as an intrinsic optimal chemical potential. It is the largest admissible value whose uncertainty interval respects $\eta\leq0.950$. The $0.10~\mathrm{mPa}$ cavity-contrast condition is likewise a design specification rather than a universal physical constant. Table~\ref{tab:tradeoff} shows that the baseline optimum is unchanged for required excesses from $0.05$ to $0.15~\mathrm{mPa}$, whereas $0.20~\mathrm{mPa}$ is infeasible in the declared box. The same table confirms the uncertainty hierarchy predicted analytically: drive uncertainty dominates the performance loss, while symmetric gap uncertainty enters more weakly.

The pressure decomposition identifies the physical mechanism. At the nominal PEC design, the propagating nonequilibrium contribution is $3.242~\mathrm{mPa}$, compared with an equilibrium contribution of $-0.1842~\mathrm{mPa}$ and an evanescent correction of $-1.210\times 10^{-3}~\mathrm{mPa}$. The positive force is therefore dominated by biased-semiconductor luminescence modified by Fabry--P\'erot scattering. The nominal cavity excess is $0.3067~\mathrm{mPa}$ and the robust cavity excess is $0.1812~\mathrm{mPa}$. For gold the corresponding robust pressure and excess remain $2.164$ and $0.160~\mathrm{mPa}$. The appropriate interpretation is cavity-enhanced nonequilibrium Casimir--Lifshitz pressure with a substantial radiation-pressure component, not a purely zero-point force.

The general far-field expression clarifies the role of the ideal mirror. The PEC maximizes the phase-averaged momentum transfer mode by mode among passive specular reflectors. This is an exact far-field statement. At finite gap, however, the PEC and gold cavity phases differ, so gold must be reoptimized rather than assigned a multiplicative conductivity correction. The gold result in Table~\ref{tab:crosscomparison} shows that the optimized lobe survives finite conductivity. The independent response check in Table~\ref{tab:gaasmodels}, whose maximum change is $0.0444\%$, further shows that this lobe is not created by fine tuning of the primary analytic GaAs fit.

The band-gap screen and thickness selection have similarly restricted meanings. The far-field screen identifies the direct-gap scale favored by the competition between photon occupation and the $E^3$ momentum-flux weighting; the subsequent finite-gap calculation uses the complete GaAs response. The value $b=13.5~\mu\mathrm{m}$ is the minimum thickness satisfying the adopted reduction indicators, not a resonance thickness that maximizes the reduced pressure. Tightening the indicator threshold increases the required thickness but does not change the independent-face pressure once the reduction is valid. At the nominal and active worst-case points, the full finite-slab calculation changes the bilateral pressure by at most
$3.62\times10^{-5}~\mathrm{mPa}$, or $1.60\times10^{-3}\%$, confirming that cross-slab transmission and correlated two-face emission are negligible at the adopted thickness.

The uniform-$\eta$ assumption still defines the principal limitation. It describes an ideal driven reservoir, while the dielectric response is held passive and independent of the source occupation. This isolates the fluctuation-force problem from carrier transport, heating, and bias-dependent optical renormalization. The gold and alternative-GaAs calculations are model-dependence tests, not a device-specific prediction. A quantitatively predictive device model would require measured optical data and a self-consistent quasi-Fermi-level profile, but those additions do not change the max--min structure established here.

\section{Conclusion}

We have formulated nonequilibrium Casimir repulsion in a reflector--vacuum--biased-GaAs--vacuum--reflector structure as a bilateral constrained max--min problem. The formulation distinguishes the outward pressures on the two boundaries, includes bounded gap and drive uncertainty, enforces a finite drive margin, selects the minimum thickness consistent with an optically thick-face reduction, and rejects designs explained almost entirely by the separation-independent far-field plateau. Within the reduced model, identical reflectors imply that a symmetric optimizer exists, so equality of the two optimal gaps is a result rather than an assumption. The generalized far-field calculation further proves that the PEC plateau is the modewise upper bound for passive specular reflectors.

For $T=300~\mathrm{K}$, $\Delta a=10~\mathrm{nm}$, and $\Delta\eta=0.005$, the PEC optimum over the declared compact box has $a_L=a_R=1.351~\mu\mathrm{m}$, $\eta=0.945$, and $b=13.5~\mu\mathrm{m}$. Its worst-case outward pressure is $2.257~\mathrm{mPa}$, with a $0.181~\mathrm{mPa}$ cavity excess. Replacing both PECs by causal Lorentz--Drude gold and reoptimizing gives $a_L=a_R=1.325~\mu\mathrm{m}$, $J^\star=2.164~\mathrm{mPa}$, and a $0.160~\mathrm{mPa}$ cavity excess. Cross-evaluation shows that robust optimization shifts either reflector design by less than $2~\mathrm{nm}$ and improves its worst-case pressure by approximately $0.0023~\mathrm{mPa}$. Its value here is therefore certification, not a claim of a large geometric redesign. The force is dominated by propagating nonequilibrium emission in both cases.

A full finite-slab evaluation changes the PEC bilateral pressure by no more than $3.62\times10^{-5}~\mathrm{mPa}$ ($1.60\times10^{-3}\%$). Replacing the primary GaAs fit with the Raki\'{c}--Majewski parameterization changes the pressure by at most $0.0444\%$ at the validation points. These two checks address different approximations: one validates the independent-face reduction, while the other probes dependence on the semiconductor response.

These results establish a reproducible robust design problem and quantify the parameter regime in which bilateral repulsion survives both admitted uncertainty and finite reflector conductivity. They do not constitute a claim of equilibrium levitation or a device-specific prediction. The main remaining physical extension is a self-consistent carrier-transport calculation of the quasi-Fermi-level profile, not a reformulation of the electromagnetic optimization.


\begin{acknowledgments}
M.R.K. gratefully acknowledges partial support from Prince Mohammad Bin Fahd University. The authors used ChatGPT to assist in drafting portions of the manuscript.
\end{acknowledgments}

\section*{Data Availability}
The numerical data underlying the main and supplemental figures, as well as the associated numerical study
, together with the Wolfram Mathematica \cite{Mathematica9} source used to generate them, are available on GitHub \cite{Khodja2026CasimirCode}.

\appendix


\appsection{Proof of the passive-reflector bound}
\label{app:farfield}

For completeness, write $\rho=|\rho|e^{\ii\theta}$ and expand
\begin{equation}
\frac{1}{|1-\rho e^{\ii\phi}|^2}
=\sum_{m=0}^{\infty}\sum_{n=0}^{\infty}
|\rho|^{m+n}e^{\ii(m-n)(\phi+\theta)}.
\end{equation}
Phase averaging gives $m=n$,
\begin{equation}
\left\langle\frac{1}{|1-\rho e^{\ii\phi}|^2}\right\rangle
=\sum_{n=0}^{\infty}|\rho|^{2n}
=\frac{1}{1-|\rho|^2}.
\end{equation}
Substitution in Eq.~\eqref{eq:ppw} gives the polarization factor
\begin{equation}
F_p=\frac{(1-|r_s^p|^2)(1+|r_c^p|^2)}
{1-|r_s^p r_c^p|^2},
\end{equation}
and therefore Eq.~\eqref{eq:pinfgeneral}. For a passive propagating channel, put $x=|r_s^p|^2$ and $y=|r_c^p|^2$. Direct differentiation gives Eq.~\eqref{eq:reflectorbound}; hence $F_p$ is nondecreasing in reflector intensity. Its limiting values are $F_p(x,0)=1-x$ and $F_p(x,1)=2$. This proves the claimed modewise PEC upper bound. In the PEC limit both polarizations give a combined factor four, and $\int_0^1t^2\dd t=1/3$ yields Eq.~\eqref{eq:pinfomega}. The equilibrium and evanescent terms vanish with increasing $a$, leaving the finite propagating plateau.

\bibliography{casimir_optimization_PRA}

\end{document}


\title{Supplemental Material for ``Passive-reflector bounds and robust bilateral nonequilibrium Casimir repulsion''}

\author{Mohamed R. Khodja}
\email{mkhodja@pmu.edu.sa}
\affiliation{Department of Electrical Engineering, Prince Mohammad Bin Fahd University, Khobar 31952, Saudi Arabia}

\author{Belkacem E. Khodja}
\affiliation{Department of Physics, Boston University, 590 Commonwealth Avenue, Boston, Massachusetts 02215, USA}

\maketitle

This document collects validation details that support, but are not needed to follow, the main argument. Numerical values are generated by the same Mathematica 9 implementation supplied with the article. Additional digits are retained in numerical-audit quantities only to display convergence and small residual differences; they should not be interpreted as corresponding physical accuracy.

\newpage 

\section{Band-gap screen}

Table~\ref{tab:materialscreenS} gives the ideal-opaque-emitter screen used before selecting a detailed material response. It is not a finite-gap calculation.

\begin{table}[h]
\caption{Ideal-opaque-emitter far-field pressure used as a band-gap screening metric. Pressures are in mPa.}
\label{tab:materialscreenS}
\begin{ruledtabular}
\begin{tabular}{lcccc}
Material & $E_g$ (eV) & $\eta=.940$ & $\eta=.945$ & $\eta=.950$\\
\hline
InSb & 0.170 & 0.145 & 0.154 & 0.163\\
InAs & 0.354 & 0.569 & 0.624 & 0.686\\
InP  & 1.344 & 2.116 & 2.762 & 3.614\\
GaAs & 1.424 & 2.076 & 2.750 & 3.650\\
ZnS  & 3.680 & 0.181 & 0.369 & 0.752\\
Ideal $E_g^\star$ & -- & 1.229 eV & 1.344 eV & 1.481 eV\\
Ideal pressure & -- & 2.139 & 2.762 & 3.658\\
\end{tabular}
\end{ruledtabular}
\end{table}

\section{Thickness-reduction indicators}

\begin{figure}[h]
\includegraphics[width=0.62\textwidth]{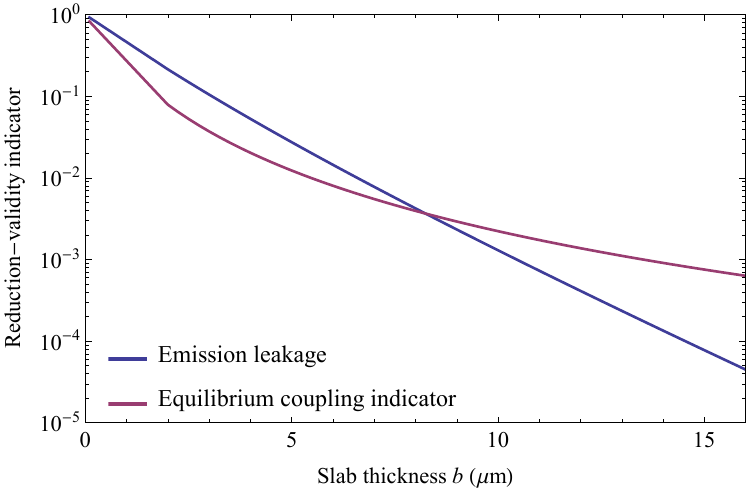}
\caption{Emission-weighted leakage and conservative equilibrium coupling indicator versus slab thickness.}
\label{fig:opacityS}
\end{figure}

Figure~\ref{fig:opacityS} shows the two prospective indicators defined in the main text. Both decrease monotonically with thickness. The adopted $b=13.5~\mu\mathrm m$ is the smallest value satisfying the two $10^{-3}$ criteria over the declared gap window.

\begin{table}[h]
\caption{Required slab thickness when the common indicator threshold is varied. Thicknesses are in $\mu$m.}
\begin{ruledtabular}
\begin{tabular}{cccc}
Threshold & Emission criterion & Equilibrium criterion & Required $b$\\
\hline
$10^{-2}$ & 6.59 & 5.46 & 6.59\\
$10^{-3}$ & 10.45 & 13.50 & 13.50\\
$10^{-4}$ & 14.53 & 30.82 & 30.82\\
\end{tabular}
\end{ruledtabular}
\end{table}

\section{Alternative GaAs response}

The optical-model check uses the Raki\'{c}--Majewski extension of Adachi's GaAs critical-point model \cite{RakicMajewski1996}. The published parameters are transcribed directly in the code: $E_0=1.410$ eV, $E_0+\Delta_0=1.746$ eV, $E_1=2.926$ eV, $E_1+\Delta_1=3.170$ eV, $A=3.97$, $\Gamma_0=0.039$ eV, $\alpha_0=1.65$, $B_1=4.15$, $B_{1s}=1.54$, $B_{1x}=1.39$, $B_{2x}=0.56$, $\Gamma_1=0.15$ eV, and $\alpha_1=0.01$. The higher critical-point terms use $(f'_0,\Gamma'_0,E'_0)=(3.89,0.55,4.49)$ eV, $(f_{2x},\Gamma_{2x},E_{2x})=(6.05,0.68,4.74)$ eV, and $(f_{2\Sigma},\Gamma_{2\Sigma},E_{2\Sigma})=(2.62,0.26,4.88)$ eV, with $\alpha_2=0.060$ and $\eps_\infty=0.77$.

The real-frequency model is used without refitting. Its absorptive part is then continued to the imaginary axis through
\begin{equation}
\eps(\ii\xi)=1+\frac{2}{\pi}\int_{0.001\,\mathrm{eV}}^{8\,\mathrm{eV}}
\frac{E\eps_2(E)}{E^2+X^2}\,\dd E
+\Delta\eps_{\rm UV}\frac{E_{\rm UV}^2}{E_{\rm UV}^2+X^2}
+\Delta\eps_{\rm ph}\frac{E_{\rm ph}^2}{E_{\rm ph}^2+X^2},
\end{equation}
where $X=\hbar\xi$ in electron-volts. The same $\eps(0)=12.9$, $\eps_\infty=10.89$, $E_{\rm UV}=10$ eV, and $E_{\rm ph}=0.035$ eV are imposed as in the primary calculation. The interband integral gives $10.1919$ at zero frequency, so $\Delta\eps_{\rm UV}=0.6981$ and $\Delta\eps_{\rm ph}=2.01$. This construction keeps the comparison causal and isolates the effect of the published real-frequency parameterization.

\section{Full finite-slab scattering formulation}

For a fixed real frequency $\omega$, transverse wavevector $k$, and polarization $p$, represent the finite GaAs slab as the reciprocal two-port scatterer
\begin{equation}
\bm S_b^p=\begin{pmatrix}\rho_p&\tau_p\\\tau_p&\rho_p\end{pmatrix}.
\end{equation}
With reference planes at its physical faces,
\begin{align}
\rho_p&=r_p\frac{1-\exp(2\ii k_{sz}b)}{1-r_p^2\exp(2\ii k_{sz}b)},\\
\tau_p&=\frac{(1-r_p^2)\exp(\ii k_{sz}b)}{1-r_p^2\exp(2\ii k_{sz}b)},
\end{align}
where
\begin{equation}
k_{sz}=\sqrt{\eps(\omega)\frac{\omega^2}{c^2}-k^2},
\qquad \operatorname{Im}k_{sz}\ge0.
\end{equation}
These expressions are analytically continued to imaginary frequency for the equilibrium term.

Let
\begin{equation}
\bm D_p=\operatorname{diag}\!\left(r_c^p e^{2\ii k_z a_L},r_c^p e^{2\ii k_z a_R}\right).
\end{equation}
If $\bm e_p$ contains the fields emitted outward by the two faces, the fields in the gaps obey
\begin{equation}
\bm b_p=\bm e_p+\bm S_b^p\bm D_p\bm b_p,
\qquad
\bm b_p=\bm U_p\bm e_p,
\qquad
\bm U_p=(\bm I-\bm S_b^p\bm D_p)^{-1}.
\end{equation}

For propagating modes the dimensionless two-face source matrix is
\begin{equation}
\bm A_p^{\rm pw}=\begin{pmatrix}A_p&B_p\\B_p^*&A_p\end{pmatrix},
\quad
A_p=1-|\rho_p|^2-|\tau_p|^2,
\quad
B_p=-\rho_p\tau_p^*-\tau_p\rho_p^*.
\end{equation}
The off-diagonal term retains correlations between radiation emitted from opposite faces. For evanescent modes, $k_z=\ii\kappa$, the corresponding matrix is
\begin{equation}
\bm A_p^{\rm ew}=\begin{pmatrix}
\operatorname{Im}\rho_p&\operatorname{Im}\tau_p\\
\operatorname{Im}\tau_p&\operatorname{Im}\rho_p
\end{pmatrix}.
\end{equation}
These are the planar scalar forms of the Rytov source-correlation operators \cite{MessinaAntezza2011,MessinaAntezza2014}; the evanescent expression is not obtained by extending $\bm I-\bm S\bm S^\dagger$ outside the propagating sector.

For $i=L,R$, the finite-slab propagating contribution is
\begin{equation}
P_{i,{\rm PW}}^{\rm fs}=\frac{\hbar}{4\pi^2}
\int_{\omega_g}^{\infty}\dd\omega\,\Delta n
\int_0^{\omega/c}k\,\dd k\,k_z\sum_p
2[\bm U_p\bm A_p^{\rm pw}\bm U_p^\dagger]_{ii}.
\end{equation}
For evanescent modes,
\begin{equation}
P_{i,{\rm EW}}^{\rm fs}=-\frac{\hbar}{\pi^2}
\int_{\omega_g}^{\infty}\dd\omega\,\Delta n
\int_{\omega/c}^{\infty}k\,\dd k\,\kappa\sum_p
r_c^p e^{-2\kappa a_i}
[\bm U_p\bm A_p^{\rm ew}\bm U_p^\dagger]_{ii},
\end{equation}
where the evanescent-sector propagation matrix is
\begin{equation}
\bm D_p=\operatorname{diag}\!\left(r_c^p e^{-2\kappa a_L},r_c^p e^{-2\kappa a_R}\right).
\end{equation}
When $\tau_p\rightarrow0$, the cross-face correlations vanish and the two expressions reduce to the independent one-gap formulas.

On the imaginary axis, the effective reflection seen from the left is
\begin{equation}
r_{{\rm eff},L}^p=\rho_p+
\frac{\tau_p^2r_c^p e^{-2\kappa a_R}}
{1-\rho_p r_c^p e^{-2\kappa a_R}},
\end{equation}
with the right expression obtained by interchanging $a_L$ and $a_R$. Hence
\begin{equation}
P_{L,{\rm eq}}^{\rm fs}=-\frac{\kb T}{\pi}
\sum_{n=0}^{\infty}{}'\int_0^\infty k\,\dd k\,\kappa_n
\sum_p\frac{r_c^p r_{{\rm eff},L}^p e^{-2\kappa_n a_L}}
{1-r_c^p r_{{\rm eff},L}^p e^{-2\kappa_n a_L}},
\end{equation}
and $P_{R,{\rm eq}}^{\rm fs}$ follows by interchange. The total is the sum of equilibrium, propagating, and evanescent terms. The comparison in the main text uses PEC amplitudes throughout this validation.

\section{Equilibrium-pressure derivative}

For one gap define $q_p(a)=R_{p,n}e^{-2\kappa_n a}$. Because $R_{p,n}$ is independent of $a$,
\begin{equation}
\frac{\partial q_p}{\partial a}=-2\kappa_nq_p,
\qquad
\frac{\partial}{\partial a}\ln(1-q_p)=\frac{2\kappa_nq_p}{1-q_p}.
\end{equation}
Inserting this derivative into $P=-\partial\mathcal F/\partial a$ gives the equilibrium pressure used in the main text. In a matrix formulation the corresponding identity is
\begin{equation}
\frac{\dd}{\dd a}\ln\det\bm M=
\operatorname{Tr}\!\left(\bm M^{-1}\frac{\dd\bm M}{\dd a}\right).
\end{equation}
Each physical propagation interval must be differentiated exactly once. Treating an effective reflection that already contains the variable gap as $a$ independent would double count or omit a propagation derivative.

\section{Continuous-uncertainty audit}

The four uncertainty corners are used only for coarse screening. At each final design, the total nonequilibrium spectral kernel is evaluated over the full admitted gap interval. The minimum ratio $K/K_{\rm PW}$ is $0.9996$ for the PEC and exceeds unity for gold. Together with $\partial_\eta\Delta n>0$, this fixes the drive minimum at $\delta\eta=-0.005$. Direct evaluation gives
\begin{equation}
\min\frac{\partial P}{\partial\eta}=
136.7~\mathrm{mPa}\quad(\mathrm{PEC}),\qquad
131.9~\mathrm{mPa}\quad(\mathrm{Au})
\end{equation}
per unit change in $\eta$ over the respective gap intervals.

The remaining one-dimensional function $P(a,\eta_0-\Delta\eta)$ is sampled on an 81-node grid. Every sign change of its centered derivative is located and refined by local quadratic interpolation, then compared with both interval endpoints. Each model has one interior stationary point, a local maximum. At the exported final designs, direct evaluation gives
\begin{align}
P_{\rm PEC}(a_0-\Delta a)&=2.256840~\mathrm{mPa},&
P_{\rm PEC}(a_0+\Delta a)&=2.256837~\mathrm{mPa},\\
P_{\rm Au}(a_0-\Delta a)&=2.164276~\mathrm{mPa},&
P_{\rm Au}(a_0+\Delta a)&=2.164276~\mathrm{mPa}.
\end{align}
The residual PEC and gold endpoint differences are
$3.16\times10^{-6}~\mathrm{mPa}$ and
$4.54\times10^{-9}~\mathrm{mPa}$, respectively, both far below the displayed precision of the headline pressures.
Nested PEC tensor grids give the values in Table~\ref{tab:gridS}. This is a numerical audit of the smooth quadrature model, not an interval-arithmetic proof.

\begin{table}[h]
\caption{Nested-grid audit of the PEC uncertainty box.}
\label{tab:gridS}
\begin{ruledtabular}
\begin{tabular}{ccccc}
$N_{\delta a}$ & $N_{\delta\eta}$ & Active $\delta a$ (nm) & Active $\delta\eta$ & $J$ (mPa)\\
\hline
21 & 11 & 10 & $-0.005$ & 2.256837\\
41 & 21 & 10 & $-0.005$ & 2.256837\\
81 & 41 & 10 & $-0.005$ & 2.256837\\
\end{tabular}
\end{ruledtabular}
\end{table}

\section{Quadrature and cutoff convergence}

Table~\ref{tab:convergenceS} varies one numerical control at a time. All other controls are held at their final values.

\begin{table}[h]
\caption{One-factor-at-a-time convergence at the nominal optimum.}
\label{tab:convergenceS}
\begin{ruledtabular}
\begin{tabular}{cccc}
Varied quantity & Low & High & Relative change in $P_{\rm tot}$\\
\hline
$N_u$ & 260 & 360 & $3.49\times10^{-15}$\\
$N_E$ & 300 & 480 & $3.66\times10^{-12}$\\
$N_{\rm PW}$ & 400 & 600 & $3.13\times10^{-10}$\\
$N_{\rm EW}$ & 300 & 480 & $0$\\
Energy window & $0.60$ eV & $0.80$ eV & $2.07\times10^{-10}$\\
$\alpha_{\max}$ & 10 & 12 & $0$\\
\end{tabular}
\end{ruledtabular}
\end{table}

\section{Asymmetry check}

\begin{figure}[h]
\includegraphics[width=0.62\textwidth]{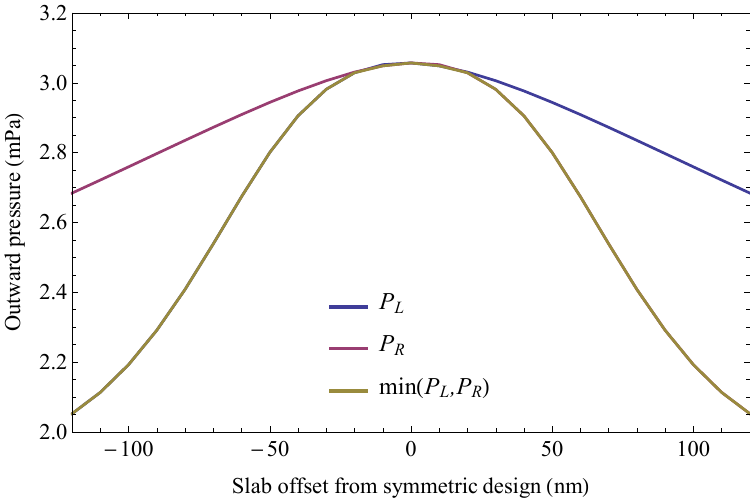}
\caption{Outward pressures and their minimum versus slab offset from the symmetric design.}
\label{fig:asymmetryS}
\end{figure}

Figure~\ref{fig:asymmetryS} moves the slab at fixed mean gap. Reflection symmetry gives $P_L(d)=P_R(-d)$. The weaker of the two pressures is maximal at $d=0$, as predicted by the bilateral symmetry argument in the main text.

\bibliography{casimir_optimization_PRA}